\documentclass[fleqn,usenatbib,useAMS]{mnras}

\usepackage[T1]{fontenc}
\usepackage{ae,aecompl}

\usepackage{bm}
\usepackage{graphicx}	
\usepackage{amsmath}	
\usepackage{amssymb}	
\usepackage{gensymb}

\usepackage{txfonts}

\title[Asteroid belt survival]{Asteroid belt survival through stellar evolution: dependence on the stellar mass} 

\author[Martin et al.]{Rebecca G. Martin\thanks{E-mail: rebecca.martin@unlv.edu}, Mario Livio, Jeremy L. Smallwood and Cheng Chen \\
Department of Physics and Astronomy, University of Nevada, Las Vegas, 4505 South Maryland Parkway, Las Vegas, NV 89154, USA
}

\date{Accepted XXX. Received YYY; in original form ZZZ}

\pubyear{2020}

\begin{document}
\label{firstpage}
\pagerange{\pageref{firstpage}--\pageref{lastpage}} 
\maketitle

\begin{abstract}
Polluted white dwarfs are {generally accreting terrestrial-like  material that may originate from a debris belt like the asteroid belt in the solar system}. {The fraction of white dwarfs that are polluted drops off significantly for white dwarfs with masses $M_{\rm WD}\gtrsim 0.8\,\rm M_\odot$}. This implies that asteroid belts and planetary systems around main-sequence stars with mass $M_{\rm MS}\gtrsim 3\,\rm M_\odot$ may not form because of the intense radiation from the star. This is in agreement with current debris disc and exoplanet observations. { The fraction of  white dwarfs  that show pollution also drops off significantly for low mass white dwarfs $(M_{\rm WD}\lesssim 0.55\,\rm M_\odot)$.} However, the low-mass white dwarfs that do show pollution are not currently accreting but have accreted in the past. We suggest that asteroid belts around  main sequence stars with masses $M_{\rm MS}\lesssim 2\,\rm M_\odot$ { are not likely to} survive the stellar evolution process. The destruction likely occurs during the AGB phase and could be the result of interactions of the asteroids with the stellar wind, the high radiation or, for the lowest mass stars  that have an unusually close-in asteroid belt, scattering during the tidal orbital decay of the inner planetary system. 
\end{abstract}

\begin{keywords} minor planets,
asteroids: general -- planets and satellites: dynamical evolution and stability -- white dwarfs -- stars: AGB and post-AGB
\end{keywords} 

\section{Introduction}
\label{intro}

White dwarfs are about $10^5$ times more dense than the Earth and so gravitational settling  of heavy elements in the atmosphere is fast, less than around a few tens of Myr \citep[e.g.][]{Paquette1986,Wyatt2014}. If the cooling age is older than this, but less than about $500\,\rm Myr$, the atmosphere consists of hydrogen and/or helium only.  Thus, the detection of metals in the atmosphere suggests accretion of material on to the white dwarf \citep[e.g.][]{Veras2016}. Observations show that at least 27\% of white dwarfs with cooling ages of $20-200\,\rm Myr$ are currently accreting debris and an additional 29\%  have accreted material in the past \citep{Koester2014}.

The composition of the white dwarf polluting material is similar to that of the bulk Earth and solar system meteorites \citep[e.g.][]{Gansicke2012,Jura2014,Xu2014,Farihi2016,Harrison2018,Hollands2018,Swan2019,Doyle2019,Bonsor2020}.
{Therefore it must have formed inside of the snow line radius, the radius outside of which water is found in the form of ice that occurs at temperatures $\sim 170\,\rm K$ in the protoplanetary disc \citep[e.g.][]{Podolak2004,Lecaretal2006,Kennedy2008,Min2011,MartinandLivio2012,MartinandLivio2013snow}.}  {The material may be delivered to the white dwarf from a planetesimal belt similar to the asteroid belt in the solar system rather than a Kuiper belt equivalent. } Only in rare cases is volatile rich material accreted \citep[e.g.,][]{Xu2017}.  {Other suggested sources include delivery by moons \citep{Payne2016,Payne2017} or fragments of broken up terrestrial planets (or moons) \citep{Malamud2020,Malamud2020b}. }
Given the observed white dwarf accretion rates it is expected that most polluted white dwarfs have a reservoir of mass at least comparable to the mass in the asteroid belt in the solar system \citep{Zuckerman2010}.

Asteroidal material is delivered to the white dwarf through a debris disc close to the white dwarf that forms through tidal disruptions \citep[e.g.][]{Jura2003,Debes2012,Veras2014,Veras2015,Xu2018,Malamud2020,Malamud2020b}. Asteroids may be perturbed into highly eccentric orbits through interactions with undetected planets \citep[e.g.][]{Debes2012,Frewen2014,Bonsor2015,Smallwood2018}. 
We suggest that for a white dwarf to be polluted over long timescales there are two requirements. First, an asteroid belt  must  form around the main-sequence star.  Secondly, the asteroid belt must survive the stellar evolution process. In Section~\ref{planetarysystems} we  examine the properties of main-sequence stars that host debris discs and planetary systems.  We further examine observational evidence for planetary systems and debris discs around evolved stars including polluted white dwarfs. In Section~\ref{connect} we propose that asteroid belts around low mass main-sequence stars {(those with mass less than about $2\,\rm M_\odot$)} are destroyed during stellar evolution leading to a lack of polluting material around low-mass white dwarfs {(those with mass less than about $0.55\,\rm M_\odot$)}.   We draw our conclusions in Section~\ref{concs}.

\section{Observations of debris discs and planetary systems}
\label{planetarysystems}

In this Section we first examine observational evidence for the formation of asteroid belts around main-sequence stars. The detection of asteroid belts themselves is difficult, but stars that host planetary systems may  have undetected asteroid belts. We consider here the range of masses of main-sequence stars that host planets and debris discs. We then explore the evidence for planetary systems around giant stars. Finally, we investigate observational evidence that suggests that asteroid belts that form around {stars with mass greater than about $2\,\rm M_\odot$} can survive through the stellar evolution process. 

\subsection{Main-sequence star systems}
\label{mainsequence}

Most stars in the Milky Way host planetary systems \citep{Cassan2012}. Nearly all observed exoplanets have been found around stars with masses $M_{\rm MS}\lesssim 3\,\rm M_\odot$.  There are only a few exceptions that have higher stellar mass. The highest mass star with a well determined mass that hosts a planet is UMa that has mass $3.09 \pm 0.07\,\rm M_\odot$ \citep{Sato2012}. {This upper mass limit is not sharp transition but a tail where the number of planets discovered decreases with host star mass \citep[e.g.][]{Reffert2015,Ghezzi2018}. Observing planets around O-type and B-type stars is difficult and so the limit is a combination of detection limitations and where planets can form and survive around more massive stars \citep{Kennedy2008,Veras2020b}.}

Debris discs are detected around about 25\% of main-sequence stars \citep{Hughes2018}. 
Discs are observed around stars with masses $\lesssim 2.4\,\rm M_\odot$ \citep{Koenig2011}. Discs around more massive stars may be photoevaporated  on timescales which are too short to be observed due to intense radiation from the host star.

{Debris discs are observed through the thermal emission of the dust and may be characterised by the infrared excess observed in their SED. The excess may generally be modelled with one or two blackbody components \citep[e.g.][]{Su2009,Su2013}. The cold components  have temperatures $<130\,\rm K$ while the warm components have temperatures $\sim 190\,\rm K$ \citep{Morales2011,Ballering2013,Chen2014}. The warm and cold components come from different radial locations with different temperatures \citep{Kennedy2014}.} Debris disc observations  show that two-component structures, like the asteroid belt and the Kuiper belt in the solar system, are common \citep[e.g.][]{Kennedy2014,Rebollido2018}. The cause for the gap between the belts is likely the formation of planets in the gap that remove all the planetesimals from the region.  \cite{Geiler2017} found that that $98\%$ of observed debris disc systems can be explained with a two-component structure rather than a one-component structure.  The few sources for which warm dust in the systems cannot be explained by this structure must be a result of cometary sources or a recent major collision or planetary system instability.

\begin{figure}
    \centering
    \includegraphics[width=\columnwidth]{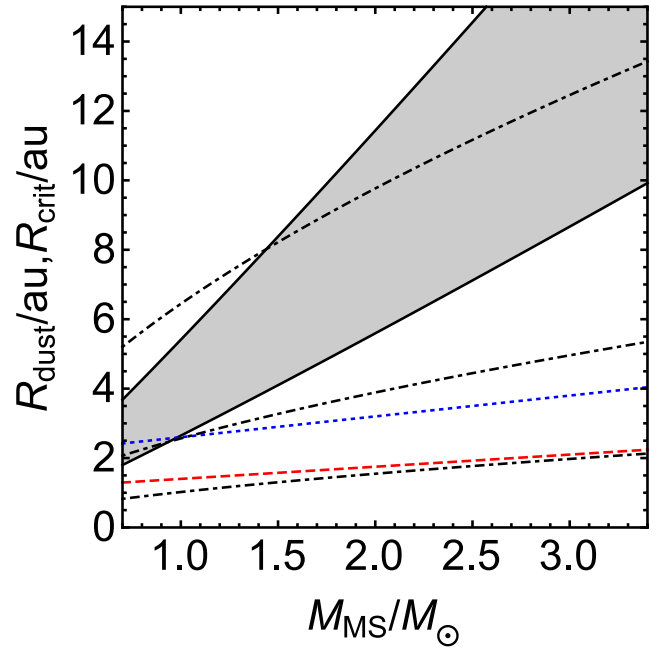}
    \caption{The shaded region shows the observed best fitting region for the location of warm dust belts (those with temperature $\sim 190\,\rm K$.) around MS stars \protect\citep[data from][]{Ballering2017}. The dotted blue line shows the critical initial semi-major axis above which a jovian planet survives the AGB phase. The dashed red line shows the critical initial semi-major axis above which a terrestrial planet survives the AGB phase. These theoretical lines are approximated from the stellar evolution models of \protect\cite{Mustill2012}. The dot-dashed lines show theoretical survival radii for $100\,\rm m$ (upper), $1\,\rm km$ (middle) and $10\,\rm km$ sized asteroids approximated by equation~(\ref{asteroid}) \citep{Dong2010}.  }  
    \label{snowline}
\end{figure}

Giant planets are thought to form outside of the snow line radius,  since  there is a higher density of solid material there \citep[e.g.][]{Pollack1996}. Thus, asteroid belts may coincide with the location of the snow line radius \citep{MartinandLivio2013asteroids}.   \cite{Ballering2017} found that the warm dust components in single-component systems (those without a cold component) are aligned with the primordial snow line, meaning the snow line in the protoplanetary disc. However, in two-component systems, the location is more diverse. The belts, at least in the one-component system, may be formed of terrestrial material. The location of the warm dust belts in one-component models have a best fit $R_{\rm dust}/{\rm au}=3.68(M/{\rm M_\odot})^{1.08}$ \citep{Ballering2017}. The shaded region in Fig.~\ref{snowline} shows the $1\sigma$ scatter around the best fitting line to the radius of warm dust belts \citep{Ballering2017}. In the two-component models the warm dust components show little correlation with stellar mass and are scattered in the approximate range $0.5-30\,\rm au$. We discuss this figure in more detail in Section~\ref{connect}.

\subsection{Giant star systems}

{To date, 112 substellar companions\footnote{https://www.lsw.uni-heidelberg.de/users/sreffert/giantplanets/giantplanets.php} around 102 G and K giant stars have been found \citep[e.g.][]{Reichert2019}. \cite{Grunblatt2019} investigated 2476 low luminosity red giant branch stars observed by the K2 mission \citep{Howell2014}. They found a higher occurence rate of planets with size greater than Jupiter in orbital periods less than 10 days compared to around dwarf Sun-like stars. This suggests that the effects of stellar evolution on the occurence of close in planets that are larger than Jupiter are not significant until the star moves significantly up the red giant branch.} 
Debris discs have also been observed around giants suggesting that they can also survive the stellar evolution \citep[e.g.][]{Bonsor2013,Bonsor2014}.  Debris discs around giant stars are more difficult to detect than around main sequence stars because radiation pressure  removes small-particle dust around higher luminosity stars \citep{Bonsor2010}.

\subsection{White dwarf systems}

The fraction of white dwarfs that host debris discs is somewhere between a few percent up to 100\%,  but most discs are too faint to detect \citep[e.g.][]{Barber2012,Veras2016,Bonsor2017,Swan2019}. The current observational limit is reached for discs around white dwarfs with cooling ages $t_{\rm cool}>0.5\,\rm Myr$ \citep{Bergfors2014}. Compact debris discs are thought to be formed by the tidal disruption of small bodies around the white dwarf \citep{Jura2003,Farihi2009,Veras2014}. Atomic emission lines suggest the existence of gaseous discs co-located with the compact circumstellar dust \citep{Gansicke2006,Guo2015}. {There is one exception to the compactness of gaseous discs, the disc around WD J0914+1914 is thought to be formed from an evaporating giant planet on a close-in orbit around the white dwarf \citep{Gansicke2019}.}

Since white dwarfs are intrinsically faint, transit searches for debris and planets are difficult. However, the light curve of WD~1145+017 shows transit features thought to be produced by dust clouds released by planetesimals that orbit the white dwarf with an orbital period of about $4.5\,\rm hr$ \citep{Vanderburg2015,Gansicke2016}. There is also evidence for solid objects orbiting around white dwarfs SDSS J1228+1040 \citep{Manser2019} {and ZTF J0139+5245 \citep{Vanderbosch2019}. The discoveries made so far have arisen from ZTF, GTC and SDSS.} \cite{vansluijs2018} examined a sample of 1148 white dwarfs observed by K2 and did not identify any substellar body transits with orbital separation $<0.5\,\rm au$.

\begin{figure}
    \centering
    \includegraphics[width=\columnwidth]{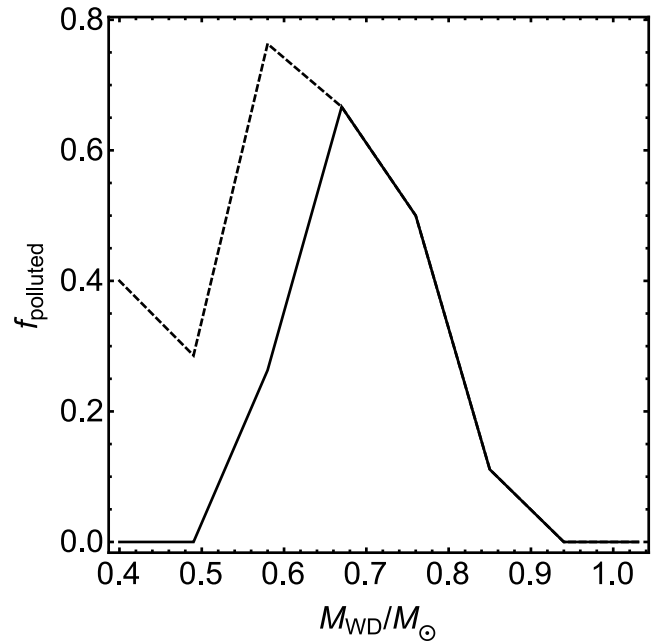}
    \caption{The fraction of the observed white dwarfs that show evidence for accretion (dashed line) and the fraction of currently accreting white dwarfs (solid line) as a function of the white dwarf mass.  The data are taken from \protect\cite{Koester2014}. } 
    \label{wddata}
\end{figure}

Fig.~\ref{wddata} shows the fraction of observed white dwarfs that are either currently accreting or show evidence for past accretion using the data from \cite{Koester2014}. { The highest mass white dwarf with evidence for pollution has mass $M_{\rm WD}=0.91\,\rm M_\odot$ \citep{Fusillo2019}. This corresponds to a progenitor main-sequence star mass of about $4\,\rm M_\odot$ \citep{Koester2014}.} {However, there is a transition where the fraction of  white dwarfs that are polluted falls off significantly at a mass of} around $M_{\rm WD}=0.8\,\rm M_\odot$.  This corresponds to a main-sequence star of around $M_{\rm MS}=3\,\rm M_\odot$. This suggests that asteroid belt formation or survival around high mass stars (those with mass $\gtrsim 3\,\rm M_\odot$) is difficult.  {Stars with mass greater than about $3\,\rm M_\odot$} are too hot for the formation of a long-lived dusty disc. This is consistent with the observations of debris discs and planetary systems around main-sequence stars discussed in Section~\ref{mainsequence}.

{Recently, \cite{Veras2020b} explored the limits on the locations of planets that would be able to survive to the white dwarf phase around stars with masses in the range $6-8\,\rm M_\odot$. They found that a major planet must be located at orbital distance greater than about $3-6\,\rm au$ at the end of the main-sequence lifetime in order to survive stellar evolution. The orbital radius outside of which minor planets survive is in the range $10-1000\,\rm au$ depending on planet size. Thus, if  white dwarf pollution is to be observed around higher mass white dwarfs in the future it would come from already fragmented debris since the minor planets would likely not be still intact. } 

While {the number of white dwarfs  included in the data drops off at low masses}, there does also appear to be a transition at small masses for which the {fraction} of  white dwarfs that are polluted drops, at around $M_{\rm WD}=0.55\,\rm M_\odot$. The white dwarfs with these low masses tend to be younger and no longer accreting. We therefore suggest that asteroid belts around low mass main-sequence stars {with mass less than $2\,\rm M_\odot$} may form, but they do not survive the stellar evolution process to the formation of the white dwarf.  We discuss possible theoretical explanations for this scenario in the next Section. 

\section{Asteroid belt destruction around low mass stars}
\label{connect}

In this Section we examine theoretical models for the evolution of asteroid orbits through stellar evolution. Our goal is to explain why the asteroid belts around stars {with mass less than about $2\,\rm M_\odot$} may not survive the process, while those around more massive stars (those with mass $2-3\,\rm M_\odot$) do.

\subsection{Planet survival}

Planets and debris that are close to the star during the main-sequence (MS) will not survive stellar evolution to the white dwarf phase as they may be engulfed or evaporated by a giant star  \citep[e.g.][]{Villaver2007,Villaver2009,Kunitomo2011}. Bodies that become engulfed by the star are expected to be destroyed unless their mass is a Jupiter mass or more \citep[e.g.][]{Livio1984,Mustill2018}. The star is largest during the AGB phase and at that time its size in AU is about equal to its initial main-sequence masses in $\rm M_\odot$ for mass in the range $1-5\,\rm M_\odot$ \citep[e.g.][]{Mustill2018}.  For higher stellar mass, there is more mass loss that occurs during the AGB phase. The mass loss leads to the expansion of the orbits of substellar bodies and therefore allows them to survive even if they begin at radii such that the stellar radius subsequently expands beyond \citep[e.g.][]{Livio1984,Mustill2012}.

There are two competing effects that determine where the critical survival orbital radius is  for a planet mass body. The tidal force pulls the object towards the expanded envelope while the effects of stellar mass loss push the planet away \citep[e.g.][]{Mustill2012}. Tidal forces are stronger for more massive planets and so the survival radius increases with planet mass. The more massive the main-sequence star, the farther out planets must be to survive engulfment. Fig.~\ref{snowline} shows approximate survival radii for initially  circular orbit terrestrial planets (dashed red line) and Jupiter mass planets (dotted blue line) \citep{Mustill2012}. The survival radii are larger for eccentric planets \citep{Villaver2014}. The corresponding lines for the RGB would be much closer in \citep[e.g.][]{Kunitomo2011, Villaver2014}.  

Terrestrial planets form inside of the snow line, and hence inside of the warm dust belt location \citep[e.g.][]{Raymond2009}. {The critical orbital radius for which terrestrial planets survive stellar evolution (the red dashed line in Fig.~\ref{snowline}) is close to the location of the warm dust belts (the shaded region in Fig.~\ref{snowline}),  } Consequently, terrestrial planets around  stars {with mass less than about $1\,\rm M_\odot$} may not survive until the white dwarf phase. However, around higher mass stars, there is a range  of orbital radii between the location of the warm dust belt and the critical survival radius where terrestrial planets may survive. 

Close-in giant planets also do not survive. 
If there is a giant planet that is engulfed, as it spirals in it may disrupt an interior asteroid belt. Thus, it would seem that for asteroid belt survival we require the giant planets to survive the engulfment process. However, as shown in Fig.~\ref{snowline}, the observed asteroid belts are at radii larger than the critical survival radius for a Jupiter mass planet.  Thus, scattering of asteroids from a planet undergoing tidal decay is unlikely, except perhaps for the lowest mass stars that have close in asteroid belts.

\subsection{Planetesimal survival}

For a planetesimal to survive stellar evolution to the white dwarf phase, it must not be engulfed by the star itself. Considering only the effects of stellar mass loss and tidal forces, this is a less stringent constraint than that which applies to  the survival of giant planets, since the planetesimal exerts only a weak tidal torque. The orbital locations of the warm dust belts are much larger than the maximum size of an AGB star (see Fig.~\ref{snowline}) and so engulfment is not likely unless there is an unusually strong  gas drag in the stellar wind.

The adiabatic approximation for the expansion of the orbits of planet and asteroid objects may be employed within about $100\,\rm au$ \citep{Veras2011,Veras2016b}. Orbital eccentricity is conserved and the relative semi-major axis increase scales with the relative stellar mass loss. Additionally, asteroids may interact strongly with the radiation from the AGB star. The interaction is complex since it depends upon the shape, orientation and albedo of each asteroid. Asteroids can be radiatively pushed by the Yarkovsky effect \citep{Bottke2001,Bottke2006,Veras2019}. The Yarkovsky drift may be several orders of magnitude larger than that from Poynting-Robertson and radiation pressure \citep{Veras2015c}. Asymmetric asteroids can be spun up through the YORP effect \citep[e.g.][]{Rubincam2000,Vokroulicky2002}. The YORP effect may destroy asteroids with sizes $100{\,\rm m}-10\,\rm km$ at orbital radii $\lesssim7\,\rm au$ \citep{Veras2014b,Veras2020}. For such conditions, the YORP affect alone may be responsible for the destruction of asteroid belts around low-mass main-sequence stars {(those with mass less than about $2\,\rm M_\odot$) },  while those around more massive stars  survive because they are at larger orbital radii. 

For asteroids with sizes small enough that the Yarkovsky effect is not important, the drag force becomes dominant \citep{Veras2015c}. The survival of planetesimals during the AGB phase may be determined by balancing the expansion of the orbit due to the stellar mass loss and the gas resistance. The critical radius for survival is 
 \begin{equation}
     R_{\rm crit}=2.57\,\left(\frac{M_{\rm MS}}{\rm M_\odot}\right)^{3/5}\left(\frac{s}{0.1\,\rm km}\right)^{-2/5}\,\rm au
     \label{asteroid}
 \end{equation}
 \citep{Dong2010}, 
where $s$ is the asteroid size and we assume a wind speed $v_{\rm wind}=10\,\rm km\,s^{-1}$ and an asteroid density of $\rho=3\,\rm g\, cm^{-3}$. In Fig.~1 we show the critical survival radius for asteroids of size $100\,\rm m$ (upper dot-dashed line), $1\,\rm km$ (middle dot-dashed line) and $10\,\rm km$ (lower dot-dashed line). The smaller asteroids in most belts may not survive the wind loss, while larger asteroids can. Asteroid belts around low-mass stars {(with mass less than about $1\,\rm M_\odot$) } may be removed for sizes $\lesssim 1\,\rm km$. Thus, asteroid belts around low mass stars may be severely depleted in mass through the interaction with the stellar wind.

\section{Conclusions}
\label{concs}

There is strong observational and theoretical evidence that white dwarf pollution  occurs from asteroid-belt-like material. {There are  significant drop offs in the fraction of  white dwarfs that are polluted at masses  higher than about $0.8\,\rm M_\odot$ and lower than about $0.55\,\rm M_\odot$}.
We have therefore proposed that (i) asteroid belts (and planetary systems) do not form around stars more massive than about $3\,\rm M_\odot$ and (ii) asteroid belts around stars less massive than about $2\,\rm M_\odot$ do not survive stellar evolution to the white dwarf stage. There are several mechanisms that can contribute to asteroid belt destruction during the AGB phase. These include the interaction of asteroids with the stellar wind through gas drag and the YORP effect, both of which affect the close-in asteroid belts around lower-mass main sequence stars. The orbital decay of a giant planet due to tides may scatter an inner asteroid belt for the very lowest mass stars.

\section*{Acknowledgments} 

We thank an anonymous referee for providing a useful and thorough review. RGM acknowledges support from NASA through grant NNX17AB96G. This research has made use of the NASA Exoplanet Archive, which is operated by the California Institute of Technology, under contract with the National Aeronautics and Space Administration under the Exoplanet Exploration Program.

\bibliographystyle{mnras} 
\bibliography{ms}
\bsp
\label{lastpage}
\end{document}